\documentclass[prd,twocolumn,showpacs,preprintnumbers,amsmath,amssymb,nofootinbib]{revtex4-1}
\usepackage{graphicx}
\usepackage{dcolumn}
\usepackage{bm}
\usepackage{epsfig}
\usepackage{dcolumn}
\usepackage{float}
\usepackage[]{hyperref}
  \hypersetup{
  unicode=false,          
  pdftoolbar=true,        
  pdfmenubar=true,        
  pdffitwindow=true,     
  pdfstartview={FitH},    
  pdfsubject={Economizing the LBNF configuration with current experiments},   
  pdfnewwindow=true,      
  pdfcreator={RevTeX},
  colorlinks=true,       
  linkcolor=red,          
  citecolor=blue,        
  urlcolor=blue,           
  }
\usepackage{hypcap}

\def\be{\begin{equation}}
\def\ee{\end{equation}}
\def\bea{\begin{eqnarray}}
\def\eea{\end{eqnarray}}\def\nn{\nonumber}
\def\gsim{\ \rlap{\raise 2pt\hbox{$>$}}{\lower 2pt \hbox{$\sim$}}\ }
\def\lsim{\ \rlap{\raise 2pt\hbox{$<$}}{\lower 2pt \hbox{$\sim$}}\ }
\def\dslash{\kern-4pt \not{\hbox{\kern-2pt $\partial$}}}
\def\pslash{\not{\hbox{\kern-2pt p}}}

\def\pmue{{{P_{\mu e}} }}

\newcommand{\nubar}{\overline{\nu}}

\newcommand{\dcp}{\delta_{CP}}
\newcommand{\nova}{NO$\nu$A}
\newcommand{\mkj}{\Delta_{21}}
\newcommand{\mlj}{\Delta_{31}}

\newcommand{\ahat}{\hat{A}}
\newcommand{\cnv}{\mbox{$\breve{\rm C}$erenkov~}}

\begin{document}
\DeclareGraphicsExtensions{.eps,.ps}
%

\title{Maximising the DUNE early 
physics output with current experiments 
}


\author{Monojit Ghosh}
\affiliation{
Physical Research Laboratory, Navrangpura,
Ahmedabad 380 009, India}
 
\author{Srubabati Goswami}
\affiliation{
Physical Research Laboratory, Navrangpura,
Ahmedabad 380 009, India}

\author{Sushant K. Raut}
\affiliation{
Physical Research Laboratory, Navrangpura,
Ahmedabad 380 009, India}
\affiliation{
Department of Theoretical Physics, School of Engineering Sciences, 
KTH Royal Institute of Technology -- AlbaNova University Center, 
Roslagstullsbacken 21, 106 91 Stockholm, Sweden}

\begin{abstract}
The Deep Underground Neutrino Experiment (DUNE)
is a  proposed next generation superbeam experiment at Fermilab. 
Its aims include measuring the unknown neutrino 
oscillation parameters -- the neutrino mass hierarchy, the octant of the 
mixing angle $\theta_{23}$ and the CP violating phase $\delta_{CP}$. 
The current and upcoming experiments T2K, \nova\ and ICAL@INO will also 
be collecting data for the same measurements. In this paper, we explore the 
sensitivity reach of DUNE in combination with these other experiments. 
We evaluate the least exposure required by DUNE to determine the 
above three unknown 
parameters with reasonable confidence. 
We find that for each case, the inclusion of data from T2K, \nova\ and 
ICAL@INO help to achieve the same sensitivity with a reduced exposure 
from DUNE thereby helping to economize the configuration. 
Further, we quantify the effect of the 
proposed near detector on systematic errors and  study the consequent 
improvement in sensitivity. 
We also examine the role 
played by the second oscillation cycle in furthering the physics reach of 
DUNE. Finally, we present an optimization study of 
the neutrino-antineutrino running of DUNE. 

\keywords{Neutrino oscillations \and DUNE \and Long-baseline}
\end{abstract}

\maketitle

\section{Introduction}
\label{intro}
The  flavour mixing of neutrinos leading to neutrino oscillations was confirmed 
by the Super-Kamiokande experiment~\cite{sk}, more than a decade ago. In the 
years since, we have measured most of the neutrino oscillation 
parameters to some precision. Solar neutrino experiments like SNO 
~\cite{Ahmad:2002jz,Aharmim:2005gt}
and the 
reactor 
neutrino experiment KamLAND~\cite{Eguchi:2002dm} have measured the
solar oscillation 
parameters $\theta_{12}$ and $\mkj$ $(=m_2^2-m_1^2)$ quite precisely. 
The atmospheric 
parameters $\theta_{23}$ and $|\mlj|$ $(=|m_3^2-m_1^2|)$ have been measured by 
Super-Kamiokande, 
MINOS and T2K ~\cite{sk_latest,minos_latest,Abe:2013fuq}. 
The smallest mixing angle $\theta_{13}$ 
has been measured 
quite recently by the reactor neutrino experiments Double Chooz, Daya Bay and 
RENO~\cite{dchooz_latest,dayabay_latest,reno_t13}. The combined fit to world 
neutrino data significantly constrains most of the oscillation parameters 
today~\cite{global_fogli,global_valle,global_nufit}.

Some quantities however, still remain unmeasured. The sign of the atmospheric 
mass-squared difference $\mlj$ is currently unknown. The case with $m_3>m_1$ $(m_3<m_1)$ 
is called Normal (Inverted) hierarchy or NH(IH). The octant in which the atmospheric 
mixing angle lies is another unknown. If $\theta_{23}<45^\circ$ $(\theta_{23}>45^\circ)$, 
then $\theta_{23}$ is said to lie in the Lower (Higher) octant or LO(HO). Finally, the 
value of the CP-violating phase $\dcp$ is completely undetermined 
with the whole range from $-180^\circ$ to $+180^\circ$ being allowed at $3\sigma$ C.L. 
However, recent hints point to a value of $\dcp$ close to 
$-90^\circ$~\cite{t2k_dcphint}.
There are many other fundamental questions like the absolute masses of the 
neutrinos and their Dirac/Majorana nature. 
However, these cannot be probed by neutrino 
oscillation experiments. 

The primary task before the current and next generation of neutrino oscillation
experiments is therefore, to measure the unknown parameters (mass hierarchy, 
octant of $\theta_{23}$ and $\dcp$) and to put more precise constraints on the 
values of the known ones. These can be achieved by experiments that probe the 
$\nu_\mu \to \nu_e$ and $\nu_\mu \to \nu_\mu$ oscillation channels at scales 
relevant to the atmospheric mass-squared difference. The superbeam  
experiments T2K~\cite{t2k} and \nova~\cite{nova} which are operational 
are the two current  
generation  long-baseline experiments that are likely 
to shed light on the above issues.  
The discovery of a non-zero $\theta_{13}$ and a precise measurement of 
this parameter 
have added a  boost to the explorations of the potential of these experiments
toward measuring the above unknowns~\cite{hubercpv,novat2k,sanjib_glade,suprabhoctant,Prakash:2013dua,octant_atmos,cpv_ino,minakata_cp,Coloma:2014kca,Ghosh:2015ena,Elevant:2015ska,Ankowski:2015kya}. 
Some of the earlier studies on this topic can be found in 
\cite{synergynt,cpcombo_sugiyama,menaparke}. 
Atmospheric neutrino experiments can also throw light on the above issues. 
One such project, ICAL@INO~\cite{inowhitepaper} is already approved and will use 
a magnetized iron calorimeter
detector with charge sensitivity.  
The combined capabilities of the long-baseline experiments T2K and 
\nova\ and the atmospheric neutrino experiment ICAL 
have been discussed extensively, e.g. see 
Refs.~\cite{octant_atmos,cpv_ino,ourlongcp,schwetzblennow,gct,Ghosh:2015ena}.  

The main problem in determining the oscillation parameters is the problem of 
parameter degeneracy~\cite{magic1,degeneracy1,degeneracy2,degeneracy3,degeneracy4,lisidegen}, 
i.e. two different sets of oscillation parameters 
giving the same value of probability. Therefore, in the degenerate parts of 
the parameter space, it is difficult for any one experiment to measure all 
the unknown parameters \cite{twobase1,twobase2,twobase3,twobase4,twobase5,twobase6}. 
Depending on the values of the oscillation parameters 
in nature, the current and upcoming experiments may be able to measure one or 
more of the unknown parameters over the next few years. 
However the expected sensitivity even in favourable parameter 
space is in the range $2 - 3 \sigma$.  
For unfavourable values of parameters as well as for enhanced 
sensitivity in the favourable region 
we will need next generation facilities. 
 The LBNE experiment~\cite{lbne} in the United States and the 
LBNO experiment~\cite{lbno_eoi} in Europe were two of the proposals for such a facility. 
Many studies have explored the physics reach of these 
experiments~\cite{laguna_options,incremental,gainfracs,suprabhlbnelbno,lbno_adequate,raj_lbne1,raj_lbne2,Dutta:2014yra,blennowstats1,Blennow:2014sja,Deepthi:2014iya,Bora:2014zwa,Ankowski:2015kya}. 
These different proposals are now converging
into a unified endeavour of a long-baseline experiment using a high-intensity beam 
from Fermilab. The proposal outlines construction of 
a deep underground neutrino observatory 
at Sanford Underground Research Facility (SURF) 
in South Dakota. This was initially called Experiment at the 
Long Baseline Neutrino Facility (ELBNF)~\cite{lbnf}, now re-christened as 
DUNE. 
The prospective detector is a modular 40 kiloton Liquid Argon 
Time Projection Chamber (LArTPC).  
One of the major goals of this facility as outlined in ~\cite{dune_cdr_v2}
is $3\sigma$ CP sensitivity for 75\% values of $\dcp$.

There are also proposals for future atmospheric neutrino experiments, 
such as HyperKamiokande~\cite{hk,hk2} which is a Water \cnv\ detector and 
PINGU~\cite{Aartsen:2014oha} which is a multi-Megaton ice detector using 
the \cnv\ technique. Some phenomenological studies involving these experiments 
have been presented in Refs.~\cite{Winter:2013ema,Choubey:2013xqa,hagiwarapingu,schwetzblennow_pingudb,akhmedovpingu}.

By the time the next generation experiments start collecting data, we will 
also have information from the current 
generation of experiments \nova, T2K and the upcoming ICAL
experiment.  
It is therefore pertinent to ask 
what the minimum amount of information 
needed from the future experiments in light of the information from this data is. 
This question was addressed in Ref.~\cite{lbno_adequate} 
in the context of the LBNO experiment. 
In that paper, three prospective baselines 
namely 2290km, 1540km and 130 km were considered for the LBNO configuration.
For the first two baselines the prototype detector 
was a LArTPC whereas for the 130 km baseline a \cnv\ detector was considered. 
It was shown that there exists a synergy between 
experiments and channels, because of which the combined analysis of many 
experiments gives very good sensitivity. Therefore the 
same physics goals can be achieved with a lower exposure
for LBNO.
In this work, we carry out a similar analysis for DUNE with a baseline of 
1300 km and taking a LArTPC detector. 
We determine the most conservative specifications that this experiment 
needs, in order to measure the remaining unknown parameters to a specified 
level of precision. 
This early physics reach of DUNE can be taken as the aim of the first 
stage, if the experiment is conducted in a staged approach.
For this purpose, we use the latest experimental specifications provided by the 
collaboration. 

In addition, 
we study the impact of the near detector (ND) in reducing  the 
systematic uncertainties,by explicitly simulating 
events at both the near detector and the far detector (FD). The role of the near 
detector and improved systematics 
used for a superbeam experiment have been considered 
in Ref.~\cite{coloma-systematic}, 
specifically in the context of the precision measurement 
of $\delta_{CP}$. 
We show the effect of systematics for all the three performance 
indicators - hierarchy, octant and $\dcp$ considering the overall 
signal and background normalization errors at both ND and FD. 

We also study the role of the second oscillation maximum 
in improving the sensitivities, both for DUNE alone and 
in conjunction with T2K, \nova\ and ICAL. 
Optimization of neutrino 
and antineutrino run has been studied before in Refs.\cite{incremental,kopphuber_2base}. 
In this work, the adequate exposure is obtained by 
assuming equal neutrino and antineutrino
runs. Subsequently we also change the proportion of neutrino and antineutrino
runs in the adequate exposure
and study what is the optimal combination. This is determined for 
each of the three unknowns for only DUNE as well as DUNE in 
conjunction with the LBL experiments T2K and \nova and the atmospheric 
neutrino experiment ICAL@INO.

The plan of the paper is as follows. 
In Section~\ref{sec:exptdet}, we discuss the configurations of the experiments considered in 
this work. The next section explores the question posed above -- determining the minimal or 
`adequate' configuration required for DUNE 
in light of data from T2K, \nova\ and ICAL, in order to determine the unknown 
parameters. We then discuss the effect of systematics 
in Section~\ref{sec:syst} and the significance of the second oscillation maximum 
for DUNE baseline in Section~\ref{sec:secondmax}. Finally, in 
Section~\ref{sec:nunubar}, we  present an optimization study of the 
neutrino-antineutrino running at  DUNE to get the best possible results.

\section{Simulation details}
\label{sec:exptdet}

Among the current generation of neutrino oscillation experiments, 
in this work we 
consider \nova, T2K and ICAL@INO. 
\nova\ and T2K are currently operational, while 
ICAL@INO project has been approved.   
The precise configuration of DUNE is still 
being worked on, and in this work we allow its specifications to be variable. 
For this work, we have simulated the long-baseline experiments using the 
GLoBES package~\cite{globes1,globes2} along with its auxiliary 
data files~\cite{messier_xsec,paschos_xsec}.

The T2K experiment in Japan shoots a beam of muon neutrinos from J-PARC to the 
Super-Kamiokande detector in Kamioka, through 295 km of earth. This experiment 
will run with a total integrated beam strength of around $8 \times 10^{21}$ pot 
(protons on target). The specifications used for this detector are as given in 
Ref.~\cite{t2k,globes_t2k3,globes_t2k4,globes_t2k5}. 
We assume in our study that T2K will run only in the neutrino mode with
the above pot. The T2K collaboration has started running in the antineutrino mode. 
For advantages of neutrino vis-\`{a}-vis antineutrino runs we refer to \cite{Ghosh:2014zea,Abe:2014tzr}. 
More discussions on the effect of antineutrino data from T2K will follow 
in the relevant sections.

The \nova\ experiment at Fermilab takes neutrinos from the NuMI beam, with a 
beam power of 0.7 MW. 
The planned run of  this experiment is for 6 years, 
divided into 3 years of neutrino  and 3 years of antineutrino mode.
The neutrinos are intercepted at the TASD detector in Ash River, 
812 km away and 14 milliradians off the beam axis. The off-axis nature of 
these  
experiments helps impose cuts to reduce the neutral current background.
 After the 
measurement of the moderately large value of $\theta_{13}$, the event selection 
criteria were re-optimized with the intention of exploiting higher 
statistics~\cite{Kyoto2012nova,sanjib_glade}. We have used this new 
configuration for the \nova\ experiment in our work. 

ICAL@INO is a magnetized iron detector for observing atmospheric neutrinos~\cite{inowhitepaper}. 
Magnetization allows for a separation of $\mu^+$ and $\mu^-$ events, and 
hence a distinction between neutrinos and antineutrinos. The total exposure 
taken for this experiment is 500 kiloton yr, i.e. 10 years of data collection 
using a 50 kiloton (kt) detector. 
We assume an energy resolution of 10\% and 
angular resolution of $10^\circ$ for the neutrinos in the detector. These 
give results comparable to the muon analysis~\cite{gct,ino2d} that 
has been performed by the INO collaboration. The new `3-d analysis' which also 
includes hadronic energy information~\cite{ino3d} is expected to give better results. 
The statistical procedure followed in calculating the sensitivity of this 
experiment follows the treatment outlined in Ref.~\cite{hri}.

DUNE is the next generation international neutrino oscillation experiment
proposed to be hosted at Fermilab. 
The beam of neutrinos, with a wide-band profile,
will travel 1300 km from Fermilab to a liquid argon detector
at SURF, South Dakota.   
The projected beam power is 1.2 MW with the possibility of upgradation to
2.4 MW. 
The detailed design of this experiment  including  beamline, detector 
and engineering aspects is expected to rely on the results of R\&D work  
already carried out by earlier proposals including LBNE and LBNO. 
The higher energy available, along with the long baseline means that the 
neutrinos will experience greater matter effects than \nova\ or T2K. 
There are two options being considered for the proton beam -- 80 GeV and 
120 GeV. For a given configuration of the beamline and 
beam power, proton energy varies inversely with the 
number of protons in the beam per unit time, and hence the neutrino flux. 
In this work, we have chosen the 120 GeV beam which gives us a 
lower flux of neutrinos and hence a conservative estimate of our results.
The details of the LArTPC detector response have been taken from 
Ref.~\cite{lbne_interim2010}. In this work we use the recently updated neutrino 
flux corresponding to 
1.2 MW beam power \cite{cherdack}. However we give our results 
in terms of MW-kt-yr.  This will enable one to interpret the results in
terms of varying detector volume, timescale and beam power. 
Note that although we use the flux corresponding to 1.2 MW beam power, if the 
accelerator geometry remains the same, then the change in the value of the 
beam power will proportionally change the flux. 
Therefore, the flux for a different value of beam power can be obtained by 
simply scaling the `standard' flux file by the appropriate factor.


Since DUNE is  proposed to be an underground observatory 
it will also be possible for it to observe atmospheric neutrinos. 
In this work, we have not considered this possibility. A detailed study on 
atmospheric neutrinos for  the DUNE experiment  is presented in
Ref.~\cite{raj_lbne1,raj_lbne2}.

The sensitivity of DUNE to the mass hierarchy, octant of $\theta_{23}$ and 
$\dcp$ comes primarily from the $\nu_\mu \to \nu_e$ oscillation 
probability $\pmue$. An approximate analytical formula for this probability can 
be derived perturbatively~\cite{cervera,freund,akhmedov} in terms of the 
two small parameters $\alpha = \mkj/\mlj$ and $\sin\theta_{13}$. 
\bea
P_{\mu e }&=
&4 \sin^2 \theta_{13} \sin^2 \theta_{23} \frac{\sin^2{[(1 -\ahat)\Delta]}}{(1-\ahat)^2}
\nn \\
&&
+ \alpha \sin{2\theta_{13}}  \sin{2\theta_{12}} \sin{2\theta_{23}} \cos{(\Delta + \dcp)}
\times \nn \\
&&
\frac{\sin{\ahat \Delta}}{\ahat} \frac{\sin{[(1-\ahat)\Delta]}}{(1-\ahat)}
+ {\cal{O}}(\alpha^2) ~.
\label{P-emu}
\eea
Here, $\Delta = \mlj L/4E$ is the oscillating term, and the effect of 
neutrinos interacting with matter in the earth is given by the 
matter term $\ahat = 2\sqrt{2} G_F n_e E/\mlj$, where $n_e$ is the 
number density of electrons in the earth. Note that this expression is valid 
in matter of constant density. This approximate formula is useful for 
understanding the physics of neutrino oscillations. However in our 
simulations, we use the full numerical probability calculated by GLoBES. 

In the analyses that follow, we have evaluated the $\chi^2$ for determining the 
mass hierarchy, the octant of $\theta_{23}$, and discovering CP 
violation {\footnote{The commonly used phrase `CP-violation discovery' 
is taken to mean the distinguishing of a given value of $\dcp$ from 
the CP-conserving cases $0$, $180^\circ$. This should not be 
confused with the statistical usage of the word `discovery' that 
implies $5\sigma$ evidence.}} using a 
combination of DUNE and the current/upcoming experiments T2K, \nova\ and ICAL. 
For each set of `true' values assumed, we evaluate the $\chi^2$ marginalized 
over the `test' parameters. In our simulations, we 
have used the effective atmospheric parameters corrected for 
three-flavour effects{ \footnote{The effective atmospheric mass-squared difference and 
atmospheric mixing angle are obtained by fitting oscillation data to the 
effective two-flavour oscillation formula. From a physics point of view, there is no 
advantage in using these effective parameters and then correcting for three-flavour effects. 
However from a computational point of view, 
we find that the use of these effective parameters gives more precise results 
while scanning the parameter space, since the hierarchy and octant degeneracies 
are exact in these parameters.
}
}~\cite{parke_defn,degouvea_defn,spuriousth23}. 
The true values assumed for the parameters are 
$\sin^2 \theta_{12} = 0.304$, $|\Delta_{31}| = 2.4 \times 10^{-3}$ eV$^2$,
$\Delta_{21} = 7.65 \times 10^{-5}$ eV$^2$ and $\sin^2 2\theta_{13} = 0.1$. 
The true value of $\dcp$ is varied throughout the full range 
$\left[ -180^\circ,180^\circ \right)$. 
For true $\theta_{23}$, we have considered three values -- $39^\circ$, $45^\circ$ 
and $51^\circ$ which are  within
the  current $3\sigma$ allowed range.
The test values of the parameters are varied in the following ranges: 
$\theta_{23} \in \left[35^\circ,55^\circ\right]$, 
$\sin^2 2\theta_{13} \in \left[0.085,0.115\right]$, 
$\dcp \in \left[-180^\circ,180^\circ\right)$. The test hierarchy is varied as well. 
The solar parameters are already measured 
quite accurately, and their variation does not impact our results significantly. 
Therefore, we have not marginalized over them.
We have imposed a prior of $\sigma(\sin^2 2\theta_{13}) = 0.005$
on the value of $\sin^2 2\theta_{13}$, which is the expected precision 
from the reactor neutrino experiments \cite{daya_005}. 
We have included backgrounds arising from NC events, mis-identified 
$\nu_\mu$ events, intrinsic beam backgrounds as well as wrong-sign 
backgrounds.
The systematic uncertainties are parameterized in terms of four nuisance 
parameters -- signal normalization error 2.5\% (7.5\%), signal tilt error 2.5\% (2.5\%), 
background normalization error 10\% (15\%) and background tilt error 2.5\% (2.5\%)~ for the 
appearance (disappearance) channel \cite{lbne}. 

\begin{figure*}[htb]
\begin{tabular}{rcl}
\epsfig{file=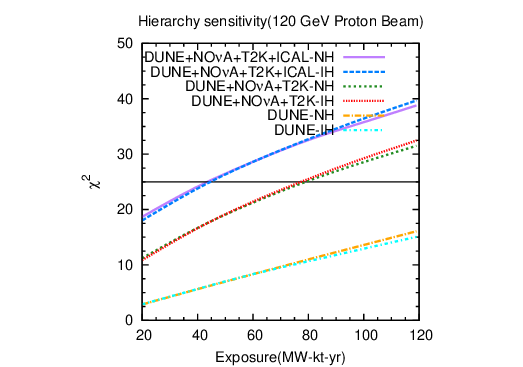, width=0.33\textwidth, bbllx=89, bblly=50, bburx=260, bbury=255,clip=}
\epsfig{file=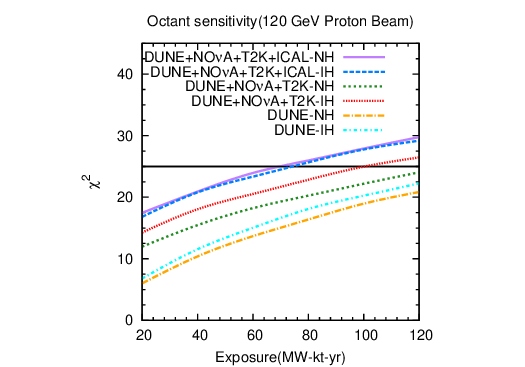, width=0.33\textwidth, bbllx=89, bblly=50, bburx=260, bbury=255,clip=}
\epsfig{file=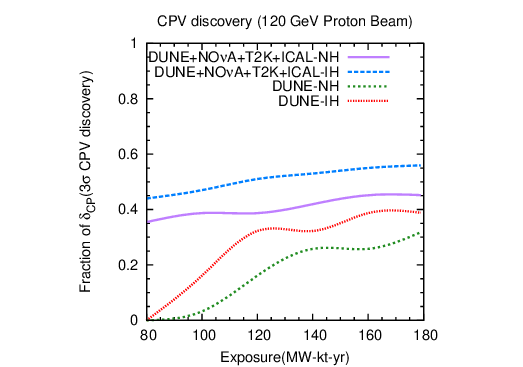, width=0.33\textwidth, bbllx=89, bblly=50, bburx=260, bbury=255,clip=}
\end{tabular}
\caption{\footnotesize Hierarchy (Octant) sensitivity $\chi^2$ vs DUNE 
exposure, for both hierarchies in the left (middle) panel. 
The value of exposure shown here is adequate to exclude the wrong hierarchy (octant) 
for all values of $\dcp$. 
Two additional sets of curves are shown to show the fall in $\chi^2$ without data
from ICAL, and the sensitivity of DUNE alone. 
The right panel shows the fraction of $\dcp$ range for which it is possible 
to exclude the CP conserving cases of $0$ and $180^\circ$, at the 
$\chi^2=9$ level.
An additional set of curves is shown to show the the CP sensitivity of DUNE alone.
}\label{fig:expo}
\end{figure*}

In section~\ref{sec:adequate}, our aim is to economize the configuration of DUNE with 
the help of the current generation of experiments. We have done that by 
evaluating the `adequate' exposure for DUNE. The qualifier `adequate', as 
defined in Ref.~\cite{lbno_adequate} in the context of LBNO, means the exposure 
required from the experiment to determine the hierarchy and octant with 
$\chi^2 = 25$, and to detect CP violation with $\chi^2 = 9$. To do so, 
we have varied the exposure of DUNE, and determined the combined 
sensitivity of DUNE along with T2K, \nova\ and ICAL. The variation of 
total sensitivity with DUNE exposure tells us what the adequate exposure 
should be. In this work, we have quantified the exposure for DUNE in units of 
MW-kt-yr. This is a product of the beam power (in MW), the runtime of 
the experiment (in years)
{\footnote{A runtime of $n$ years is to be interpreted as $n/2$ years each in 
neutrino and antineutrino mode. In this work, we have always considered equal 
runs in both modes for DUNE unless otherwise mentioned.}}
and the detector mass (in kilotons). As a 
phenomenological study, we will only specify the total exposure in this paper. 
This may be interpreted experimentally as different combinations of 
beam power, runtime and detector mass whose product quantifies the  
exposure. For example, an exposure of 40 MW-kt-yr could be achieved by 
using a 10 kt detector for 2 years (in each, $\nu$ and $\overline{\nu}$ mode), 
with a 1 MW beam.
We use events in the energy range 0.5 - 10 GeV for DUNE which covers both 
first and second oscillation maxima. The  relative contribution 
of the second oscillation maximum is discussed in Section~\ref{sec:secondmax}. 

\section{Adequate exposure for DUNE}
\label{sec:adequate}

\subsection{Hierarchy sensitivity}

In the left panel of Fig.~\ref{fig:expo}, we have shown the combined sensitivity of DUNE, 
\nova, T2K and ICAL for determining the mass hierarchy, as the exposure for 
DUNE is varied. The hierarchy sensitivity typically depends very strongly on 
the true value of $\dcp$ and $\theta_{23}$. In this work, we 
are interested in finding out the least exposure needed for DUNE, 
irrespective of the true values of the parameters in nature. Therefore, we 
have evaluated the $\chi^2$ for various true values of these parameters 
as listed in Section~\ref{sec:exptdet}, and taken the most conservative 
case out of them. Thus, the exposure plotted here is for the most 
unfavourable values of true $\dcp$ and $\theta_{23}$. Since hierarchy 
sensitivity of the $\pmue$ channel increases with $\theta_{23}$, the 
worst case is usually found at the lowest value considered -- 
$\theta_{23} = 39^\circ$. The most unfavourable of $\dcp$ is around 
$+(-)90^\circ$ for NH(IH) \cite{novat2k}. 
Separate curves are shown for both hierarchies, but the results 
are almost the same in both cases. 
We find that 
the adequate exposure for DUNE including T2K, \nova\ and ICAL data 
is around 44 MW-kt-yr 
for both NH and IH. 
This is shown by the upper curves. 
The two intermediate curves show the same sensitivity, but without including 
ICAL data in the analysis. In this case, the adequate exposure is around 
78 MW-kt-yr. 
Thus, in the absence of ICAL data, DUNE would have to increase 
its exposure by over 75\% to achieve the same results. 
For the benchmark values of 1.2 MW power and 10 kt 
detector, the exposure of 44 MW-kt-yr  implies 
under 2 years of running in each mode
whereas 
the adequate exposure
78 MW-kt-yr 
corresponds to about 3 years exposure in each mode.  

Finally, we show 
the sensitivity from DUNE alone, in the lower most curves. For the range 
of exposures considered, DUNE can achieve hierarchy sensitivity up to 
the $\chi^2=16$ level.    
The first row of Table ~\ref{tab:adequate} shows the adequate exposure 
required for hierarchy sensitivity reaching $\chi^2 = 25$ for 
only DUNE and also  
after adding the data from T2K, \nova\ and ICAL. 
The numbers in the parentheses correspond to IH. With only DUNE, the 
exposure required to reach $\chi^2=25$ for the hierarchy sensitivity is seen to be 
much higher.

\begin{table*}
\begin{center}
\begin{tabular}{|c|c|c|c|}
\hline
Sensitivity   & DUNE+NO$\nu$A+T2K+ICAL  & DUNE+NO$\nu$A+T2K & DUNE   \\          
\hline
Hierarchy($\chi^2=25$)         & 44(44)    &  78(78)  & 190(212)    \\
Octant($\chi^2=25$)            & 74(74)     &  130(100)  & 168(152)    \\
CP(40$\%$ at $\chi^2=9$)        & 130(72)     &  130(72)  & 228(180)    \\
\hline

\hline
\end{tabular}
\end{center}
\caption{Adequate exposures for hierarchy, octant and CP in units of MW-kt-yr for NH(IH)  }
\label{tab:adequate} 
\end{table*}

\begin{figure*}[htb]
\begin{tabular}{rcl}
\epsfig{file=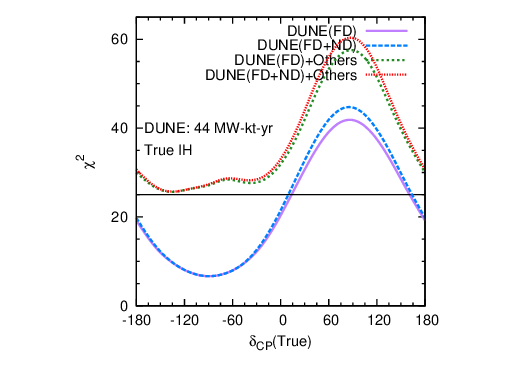, width=0.33\textwidth, bbllx=89, bblly=50, bburx=260, bbury=255,clip=}
\epsfig{file=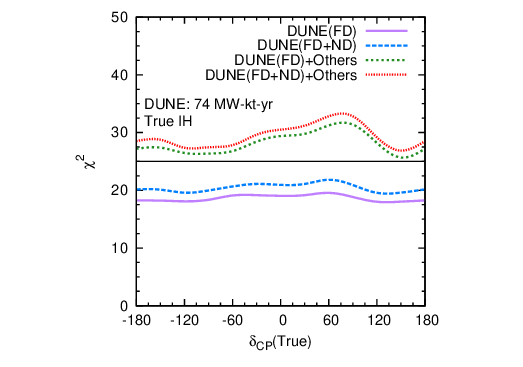, width=0.33\textwidth, bbllx=89, bblly=50, bburx=260, bbury=255,clip=}
\epsfig{file=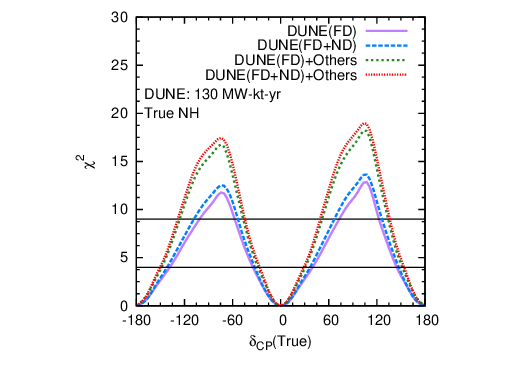, width=0.33\textwidth, bbllx=89, bblly=50, bburx=260, bbury=255,clip=}
\end{tabular}
\caption{\footnotesize Hierarchy/Octant/CP violation discovery sensitivity 
$\chi^2$ vs true $\dcp$ in the left/middle/right panel. The various curves show 
the effect of including a near detector on the sensitivity of 
DUNE alone and DUNE combined with the other experiments. 
}\label{fig:syst}
\end{figure*}

\subsection{Octant sensitivity}

The  mass hierarchy as well as the values of 
$\dcp$ and $\theta_{23}$ in nature affect  the octant sensitivity
of experiments significantly. In our analysis, we have considered various 
true values of $\dcp$ across its full range, and two representative 
true values 
of $\theta_{23}$ -- $39^\circ$ and $51^\circ$. 
Having evaluated the 
minimum $\chi^2$ for each of these cases, we have chosen the lower 
value. Thus, we have ensured that the adequate exposure
shown here holds, irrespective of the true octant of $\theta_{23}$. 
Note that octant sensitivity reduces as we go more toward $\theta_{23}= 45^\circ$. 
Thus the above choice of true $\theta_{23}$ only corresponds to the more 
conservative value of $\theta_{23}$ out of $39^\circ$ and $51^\circ$.

The middle panel of Fig.~\ref{fig:expo} shows the combined octant 
sensitivity of the experiments, as a function of DUNE exposure. 
Around  70-74 MW-kt-yr for NH(IH) is the required exposure 
for DUNE, to  measure the octant 
with 
 \nova, T2K and ICAL
This implies a runtime of around 3 years in 
each mode for the `standard' configuration of DUNE. 
Without 
information from ICAL however, DUNE would have to increase its exposure to 
around 130(100) MW-kt-yr for NH(IH) to measure the octant with $\chi^2=25$.
For a 1.2 MW beam and a 10 kt detector this implies about 
5(4)  years  for NH (IH) in each mode.  
DUNE-only would need a higher exposure of 168 (152) MW-kt-yr 
for NH(IH) corresponding to about 7(6) years in each mode. 
Thus including ICAL data reduces the exposure required from DUNE. 
This is summarized in the second row of Table ~\ref{tab:adequate}.

\subsection{Detecting CP violation}
The CP detection ability of an experiment is defined as its ability to 
distinguish the true value of $\dcp$ in nature from the CP-conserving 
cases of $0$ and $180^\circ$. This obviously depends on the true value of 
$\dcp$. If $\dcp$ in nature is close to $0$ or $180^\circ$, this 
ability will be poor, while if it is close to $\pm 90^\circ$, it will be 
high. CP detection also depends on $\theta_{23}$, and 
typically it is
a decreasing function of $\theta_{23}$ \cite{ourlongcp}. 
Here, we have tried to 
determine the fraction of the entire $\dcp$ range for which our 
setups can detect CP violation with at least $\chi^2=9$. We have always 
chosen the smallest fraction over various values of $\theta_{23}$ 
($39^\circ$, $45^\circ$ and $51^\circ$), 
so as to get a conservative estimate. 

We find in the right panel of Fig.~\ref{fig:expo} that for the range of 
exposures considered, 
the fraction of $\dcp$ is between 0.35 and 0.55. While the exposure 
increases by a factor of 2, the increase in the fraction of $\dcp$ is 
very slow. In Ref.~\cite{cpv_ino}, it was shown that the addition of
information 
from ICAL to \nova\ and T2K increases their CP detection ability. This is 
because ICAL data breaks the hierarchy-$\dcp$ degeneracy that \nova\ and T2K 
suffer from. However, the DUNE experiment itself is also capable of lifting this 
degeneracy 
for most of the values of $\dcp$ \cite{suprabhlbnelbno}. 
Therefore, the inclusion of ICAL data does not make any difference 
in this case. This combination of experiments can detect CP violation over 
40\% of the $\dcp$ range with an exposure of about 130 MW-kt-yr at DUNE for NH 
(i.e. a runtime of around 5.5 years in each mode for DUNE with the initial 10 kt detector 
or around 1.5 years in each mode with the final 40 kt detector). 
Without including T2K and \nova\ information the exposure required will 
be 228 MW-kt-yr for 40\% coverage for discovery of $\dcp$. 
As mentioned in the introduction, one of the mandates of  DUNE is
$3\sigma$ CP coverage for 75\% values of $\dcp$ ~\cite{dune_cdr_v2}. 
We find that 
an exposure of 300 MW-kt-yr in neutrinos and 300 MW-kt-yr in antineutrinos
gives 69\%(73\%) CP coverage at
$3 \sigma$ for $\theta_{23}=39^\circ$ and 60\%(65\%) for $51^\circ$ in NH(IH). We also find that
addition of NO$\nu$A and T2K data does not help much for such
high values of exposure. The results
are summarized in Table ~\ref{tab:lbnf_frac}. 

\begin{table*}
\begin{center}
\begin{tabular}{|c|c|c|}
\hline
3$\sigma$ CPV coverage for $\theta_{23}$    &  DUNE   & DUNE+NO$\nu$A+T2K   \\          
\hline
$39^o$         & 69(73)    &  71(74)    \\
$51^o$            & 60(65)     &  63(67)    \\
\hline

\hline
\end{tabular}
\end{center}
\caption{CPV coverage fraction at 3$\sigma$ for total 600 MW-Kt-yr exposure}
\label{tab:lbnf_frac} 
\end{table*}

In the following sections, we fix the exposure in each case to be the 
adequate exposure as listed in Table ~\ref{tab:adequate}, for the 
most conservative parameter values. 

\section{Role of the Near Detector in reducing systematics}
\label{sec:syst}

The measurement of a relatively large value of $\theta_{13}$ 
makes the issue of systematic uncertainties more relevant. 
The role of the ND in long-baseline neutrino experiments 
has been well discussed in the literature; see for example 
Refs.~\cite{nd_t2k1,nd_t2k2,nd_minos}. The measurement of events at the 
ND and FD reduces the uncertainty associated with the 
flux and cross-section of neutrinos. Thus the role of the near 
detector is to reduce systematic errors in the 
oscillation experiment.  
It has recently been found that the ND for 
the T2K experiment can bring about a spectacular reduction of systematic 
errors \cite{nd_t2k3}. 
The impact of ND and 
systematic uncertainties in the context of a measurement
of CP violation using appearance channels has been studied in \cite{oldsyst_schwetz,coloma-systematic}
taking the T2HK experiment as an example. 

In this study, we have tried to quantify the improvement in results, once 
the ND is included. 
The conventional way of doing this is to assume 
that the existence of the ND leads to a reduction of 
systematic effects, and therefore input smaller systematic errors
by hand in the analysis.
Instead of using this approach, we have explicitly 
simulated the events at the ND using 
GLoBES. The design for the ND is still being planned. For our 
simulations, we assume that the ND has a mass of 5 tons and 
is placed 459 meters from the source. The flux at the ND site 
has been provided by the DUNE collaboration \cite{cherdack}. The 
detector characteristics for the ND are as follows \cite{nd_lbneindia}. 
The muon(electron) detection efficiency is taken to be 95\%(50\%). The NC 
background can be rejected with an efficiency of 20\%. The energy resolution 
for electrons is $6\%/ \sqrt{E \rm{(GeV)} }$, while 
that for muons is $37$ MeV across the entire energy range of interest. 
Therefore, for the neutrinos, we use a (somewhat conservative) energy 
resolution of $20\%/ \sqrt{E \rm{(GeV)} }$. The systematic errors that the 
ND setup suffers from are assumed to be the same as those from 
the FD. 

In order to have equal runtime for both FD and ND, we fix
the FD volume as 10 kt and consider both the detectors to 
receive neutrinos from 1.2 MW beam. This
fixes the runtime of FD which is then also used in the simulation for ND. 
The run times used in this section are chosen corresponding to
the adequate exposures from the previous section as given in the first column of Table \ref{tab:adequate}:
3.6 year for hierarchy sensitivity, 6.2 year for octant sensitivity and 
10.8 year for CPV discovery sensitivity.

In order to simulate the ND+FD setup for DUNE, we use GLoBES to generate 
events at both detectors, treating them as separate experiments. 
We then use these two data sets to perform a correlated systematics analysis using 
the method of pulls \cite{pulls_gg}. This gives us the combined sensitivity 
of DUNE using 
both ND and FD. (We have explained our methodology in 
\ref{app:syst}.) Thereafter, the procedure of combining results 
with other experiments and marginalizing over oscillation parameters 
continues in the usual manner. The results are shown in 
Fig.~\ref{fig:syst}. The effect of reduced systematic errors is felt most 
significantly in regions where the results are best. This is because 
for those values of $\dcp$, the experiment typically has high enough 
statistics for systematic errors to play an important role. 

\begin{figure}[htb]
\begin{center}
\epsfig{file=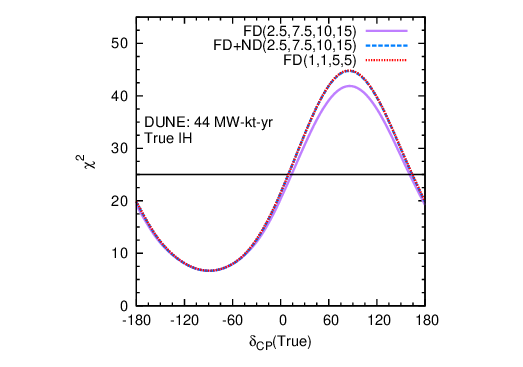, width=0.4\textwidth, bbllx=80, bblly=50, bburx=275, bbury=220,clip=}
\end{center}
\caption{\footnotesize Reduction in systematics due to inclusion of the 
near detector. The numbers in brackets denote $\nu_e$ appearance signal 
normalization error, $\nu_\mu$ disappearance signal normalization error, $\nu_e$ appearance 
background normalization error and $\nu_\mu$ disappearance background normalization error.
}
\label{fig:syst2}
\end{figure}

Next, we have tried to quantify the reduction in systematic errors seen 
by the experiment, when the ND is included. To be more 
specific, if the systematic errors seen by each detector setup are
denoted by 
$\vec{\pi}$, then we wonder what is the effective set of errors 
$\vec{\pi}_{\rm eff}$ for the FD setup, once the 
ND is also included. In other words, for given systematic 
errors $\vec{\pi}$, we have found the effective errors $\vec{\pi}_{\rm eff}$ 
that satisfy the relation
\begin{equation}
 \chi^2 ({\rm{FD}} (\vec{\pi}_{\rm eff})) \equiv 
 \chi^2 ({\rm{FD}} (\vec{\pi}) \oplus {\rm{ND}} (\vec{\pi}) ) ~,
\end{equation}
where the right-hand side denotes the correlated combination as described 
in ~\ref{app:syst}. The $\vec{\pi}_{\rm eff}$ thus computed 
can be used in future simulations as the reduced set of systematic errors 
because of the presence of the ND. We have 
chosen typical values of systematic errors for the detector: 
$\nu_e$ appearance signal normalization error of 2.5\%, $\nu_\mu$ disappearance signal 
normalization error of 7.5\%, $\nu_e$ appearance background normalization error of 10\% and 
$\nu_\mu$ disappearance background normalization error of 15\%.
The tilt error is taken as 2.5\% in both appearance and 
disappearance channels.
The first four numbers 
constitute $\vec{\pi}$, as labeled in the 
figure.
We find that the tilt errors have a very small effect in this particular 
analysis, and we fix them to the value specified above.
The result of the computation is shown in 
Fig.~\ref{fig:syst2}, for the case of hierarchy determination. The sensitivity 
of FD+ND obtained using these numbers, are matched by an FD setup with 
effective errors as follows: $\nu_e$ appearance signal normalization error of 1\%, 
$\nu_\mu$ disappearance signal 
normalization error of 1\%, $\nu_e$ appearance background normalization error of 5\% and 
$\nu_\mu$ disappearance background normalization error of 5\%. Similar results 
are obtained in the case of octant and CP sensitivity also. Thus, inclusion 
of the ND brings the systematic errors down to 13-50\% of their 
original value. 
These results are summarized in Table~\ref{tab:systresults}. 
Note that the numbers presented in Table ~\ref{tab:systresults} 
are indicative assuming the systematic uncertainties are energy independent. 
However, the improvement in the systematic uncertainties 
in the actual analysis incorporate this energy dependence due to a full 
bin-by-bin analysis of the ND data.

\begin{table}[htb]
  \begin{tabular}{|l|c|c|}
  \hline
  Systematic error & only FD & FD+ND \\
  \hline
   $\nu_e$ app signal norm error & 2.5\% & 1\% \\
   $\nu_\mu$ disapp signal norm error & 7.5\% & 1\% \\
   $\nu_e$ app background norm error & 10\% & 5\% \\ 
$\nu_\mu$ disapp background norm error & 15\% & 5\% \\
\hline
  \end{tabular}
\caption{\footnotesize Reduction in systematic errors with the addition 
of a near detector
}
\label{tab:systresults}
\end{table}

\section{Significance of the second oscillation maximum}
\label{sec:secondmax}

For a baseline of 1300 km, the oscillation probability $P_{\mu e}$ has its 
first oscillation maximum around 2-2.5 GeV. This is easy to explain from 
the formula 
\[
 \frac{\mlj^{(m)} L}{4 E} = \frac{\pi}{2} ~,
\]
where $\mlj^{(m)}$ is the matter-modified atmospheric mass-squared difference. 
In the limit $\mkj \to 0$, it is given by 
\[
 \mlj^{(m)} = \mlj \sqrt{ (1-\hat{A})^2 + \sin^2 2\theta_{13} } ~.
\]
The second oscillation maximum, for which the oscillating term takes the 
value $3\pi/2$, occurs at an energy of around 0.6-1.0. Studies have 
discussed the advantages of using the second oscillation maximum to 
get information on the oscillation 
parameters~\cite{wbb_vs_oa,kopphuber_2base}. 
In fact, one of the main aims of the proposed ESSnuSB 
project~\cite{ess,suprabh_ess} is to study neutrino oscillations 
at the second oscillation maximum. 

\begin{figure}[htb]
\begin{center}
\epsfig{file=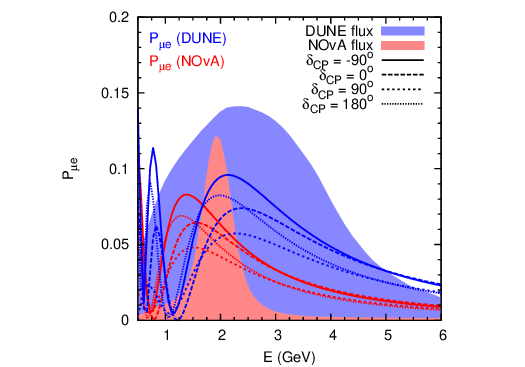, width=0.4\textwidth, bbllx=89, bblly=50, bburx=275, bbury=220,clip=}
\end{center}
\caption{\footnotesize Neutrino oscillation probability $P_{\mu e}$ for 
various representative values of $\dcp$ and normal hierarchy, 
for the \nova\ and DUNE baselines. 
Also shown as shaded profiles in the background are the $\nu_\mu$ flux
for both these experiments (on independent, arbitrary scales).
}
\label{fig:beamprofile}
\end{figure}

The neutrino 
flux that DUNE will use has a wide-band profile, which can extract 
physics from both, the first and second maxima. Figure~\ref{fig:beamprofile} 
shows $P_{\mu e}$ for the DUNE baseline, superimposed on the $\nu_\mu$ flux. 
This is in contrast with \nova, which uses a narrow-band 
off-axis beam concentrating on its first oscillation maximum, in order 
to reduce the $\pi^0$ background at higher energies.

\begin{figure*}[htb]
\begin{tabular}{rcl}
\epsfig{file=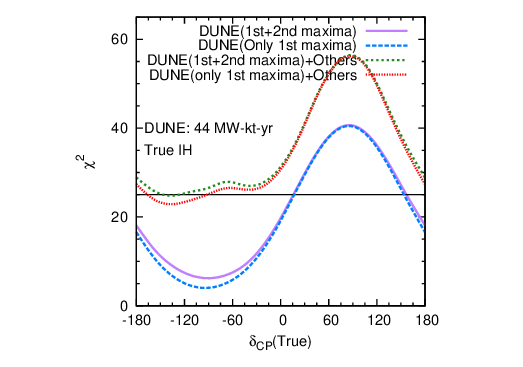, width=0.33\textwidth, bbllx=89, bblly=50, bburx=260, bbury=255,clip=}
\epsfig{file=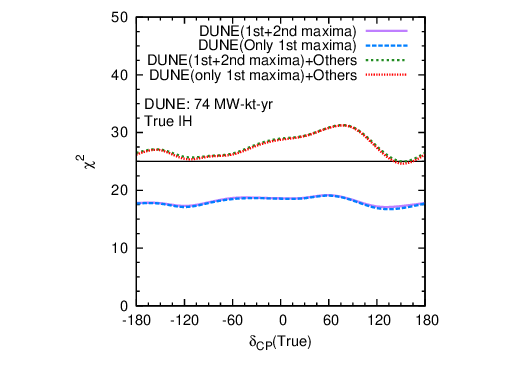, width=0.33\textwidth, bbllx=89, bblly=50, bburx=260, bbury=255,clip=}
\epsfig{file=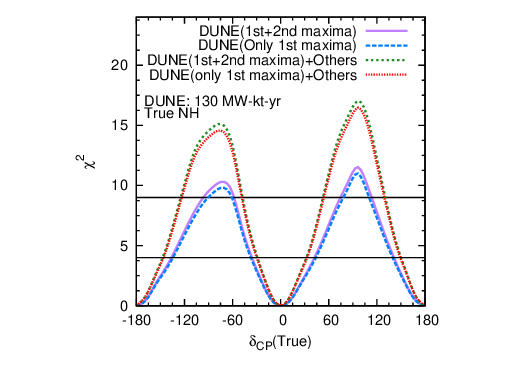, width=0.33\textwidth, bbllx=89, bblly=50, bburx=260, bbury=255,clip=}
\end{tabular}
\caption{\footnotesize Hierarchy/Octant/CP violation discovery sensitivity 
$\chi^2$ vs true $\dcp$ in the left/middle/right panel. The various curves show 
the effect of data from the second oscillation maximum on the sensitivity of 
DUNE alone and DUNE combined with the other experiments. 
}\label{fig:secondmax}
\end{figure*}

\begin{figure*}[htb]
\begin{center}
\begin{tabular}{cc}
\epsfig{file=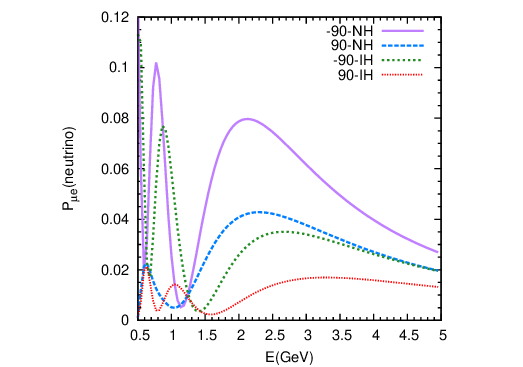, width=0.33\textwidth, bbllx=89, bblly=50, bburx=280, bbury=255,clip=}
\epsfig{file=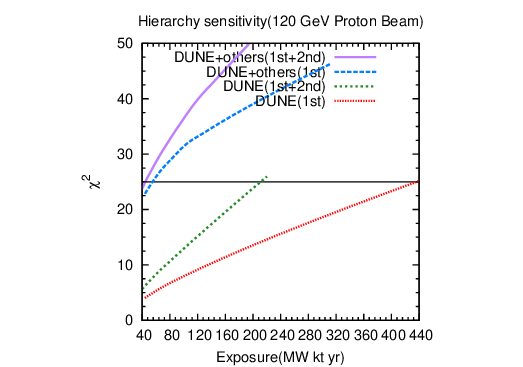, width=0.33\textwidth, bbllx=89, bblly=50, bburx=260, bbury=255,clip=}
\end{tabular}
\end{center}
\caption{\footnotesize Left panel: The probability vs energy showing 
the hierarchy-$\dcp$ degeneracy affecting hierarchy sensitivity at the 
two oscillation maxima. Right panel: Hierarchy exclusion
$\chi^2$ vs exposure, with and without the second oscillation maximum. 
}
\label{fig:adequate2nd}
\end{figure*}

In order to understand the impact of the second oscillation maximum, 
we have considered two different energy ranges. Above 1.1 GeV, only 
the first oscillation cycle is relevant. However, if we also include 
the energy range from 0.5 to 1.1 GeV, we also get information from 
the second oscillation maximum. Figure~\ref{fig:secondmax} compares the 
sensitivity to the hierarchy, octant and CP violation only 
from the first oscillation cycle, and from both the oscillation cycles
assuming the adequate exposures obtained in the previous section.
We see 
that inclusion of data from the second oscillation maximum only increases 
the $\chi^2$ by a small amount. This increase is visible only 
for hierarchy sensitivity.  The effect is  seen to be more pronounced in 
the region $\dcp \sim -90^\circ$.
This is because the probability for \{IH,$\dcp = -90^\circ$\}  
is closer to that for \{NH,$\dcp = +90^\circ$\} at the first oscillation
maximum, as reflected in the first panel of Fig.~\ref{fig:adequate2nd}. 
But at the second oscillation maximum the separation between the
probabilities for these two sets is higher. Therefore adding the 
second oscillation maximum aids the hierarchy sensitivity.  

In the right panel of Fig.~\ref{fig:adequate2nd} we  show how the exposure 
for hierarchy sensitivity depends on the inclusion of the 
second oscillation maximum for only DUNE and DUNE+T2K+\nova+ICAL. 
It is clear from the figure that the second maximum plays a more significant 
role for higher exposure. 
For the combined case, $5\sigma$ sensitivity is reached at a relatively 
lower exposure and hence the second maximum does not play a major role. 
This is also seen in the figure \ref{fig:secondmax}. 
However only for DUNE since $5\sigma$ sensitivity is reached for 
a relatively higher exposure the inclusion of the second oscillation 
maximum is seen to play an important role.  
This feature is reflected in Table~\ref{tab:secondmax}.

\begin{table*}[htb]
\begin{tabular}{|c|c|c|c|c|}
\hline
  Sensitivity   & \multicolumn{2}{|c|}{DUNE+NO$\nu$A+T2K+ICAL(MW-kt-yr)} & \multicolumn{2}{|c|}{Only DUNE(MW-kt-yr)} \\
\cline{2-5}
                & 1st + 2nd osc. cycles & Only 1st osc. cycle & 1st + 2nd osc. cycles & Only 1st osc. cycle \\
\hline
Hierarchy ($\chi^2 = 25$)                  & 44       &  56   & 212  &  436  \\
Octant ($\chi^2 = 25$)                     & 74         &  78           & 168   &  190   \\
CP ($40\%$ coverage at $\chi^2 = 9$)       & 130        &  140           & 228  &  256  \\
\hline
\end{tabular}
\caption{Effect of the second oscillation maximum on the sensitivity of DUNE. The 
numbers indicate the adequate exposure (in MW-kt-yr) required by DUNE for determining 
the oscillation parameters, with and without the contribution from the second 
oscillation maximum. For each of the three unknowns, the true parameters 
(including hierarchy) are taken 
to be ones for which we get the most conservative sensitivity.}
\label{tab:secondmax}
\end{table*}

\section{Optimizing the neutrino-antineutrino runs}
\label{sec:nunubar}

One of the main questions while planning any beam-based neutrino experiment 
is the ratio of neutrino to antineutrino run. Since the dependence 
of the oscillation parameters on the neutrino and antineutrino  
probabilities are different, an antineutrino run can provide a different set of 
data which may be useful in determination of the parameters. However, the 
interaction cross-section for antineutrinos in the
detectors is smaller by a factor 
of 2.5-3 than the neutrino cross-sections. Therefore, an antineutrino run typically 
has lower statistics. Thus, the choice of neutrino-antineutrino ratio is often 
a compromise between new information and statistics. 

\begin{figure*}
\begin{tabular}{cc}

 \epsfig{file=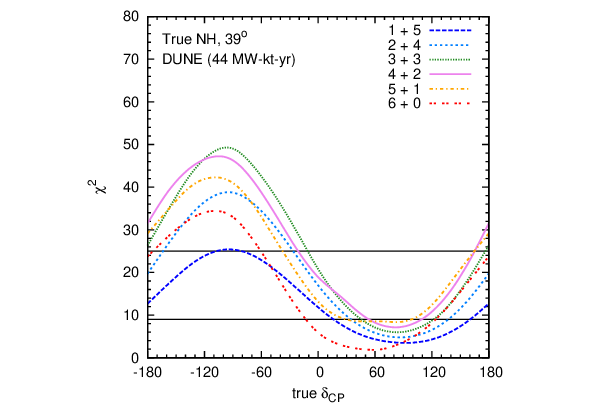, width=0.40\textwidth, bbllx=89, bblly=50, bburx=300, bbury=255,clip=}&
 \epsfig{file=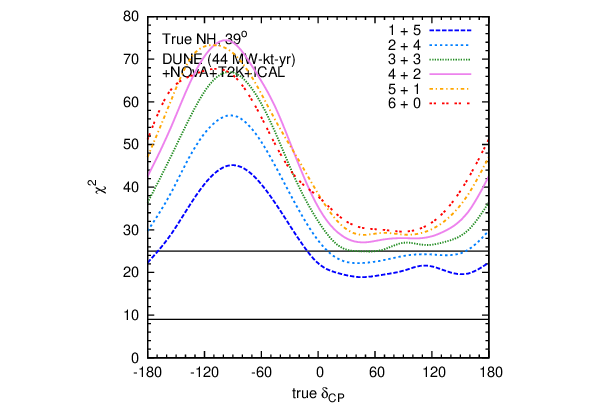, width=0.40\textwidth, bbllx=89, bblly=50, bburx=300, bbury=255,clip=}
 \\
 \epsfig{file=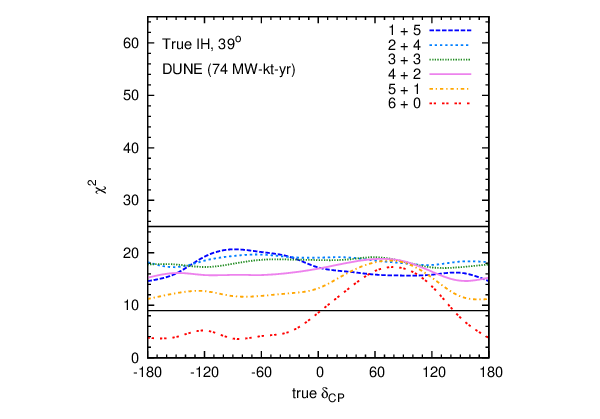, width=0.40\textwidth, bbllx=89, bblly=50, bburx=300, bbury=255,clip=}&
 \epsfig{file=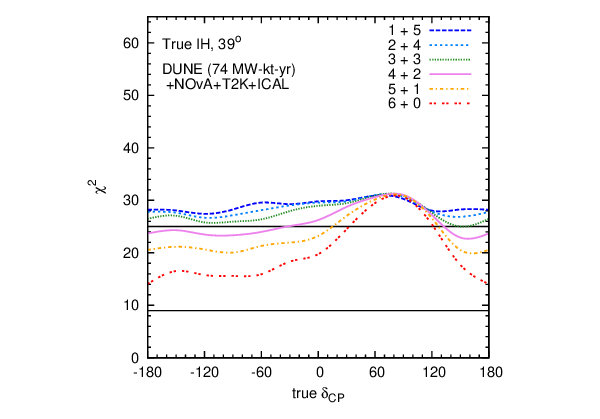, width=0.40\textwidth, bbllx=89, bblly=50, bburx=300, bbury=255,clip=}
 \\
 \epsfig{file=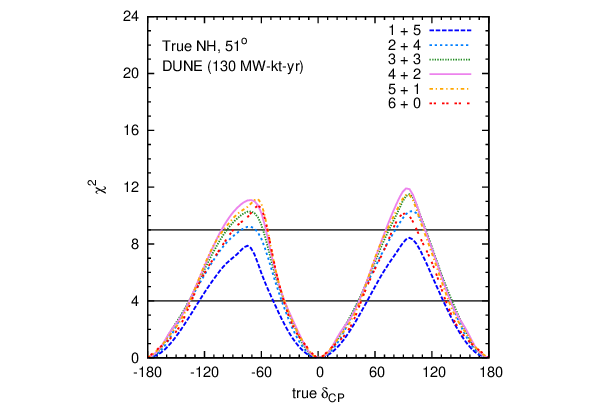, width=0.40\textwidth, bbllx=89, bblly=50, bburx=300, bbury=255,clip=}&
 \epsfig{file=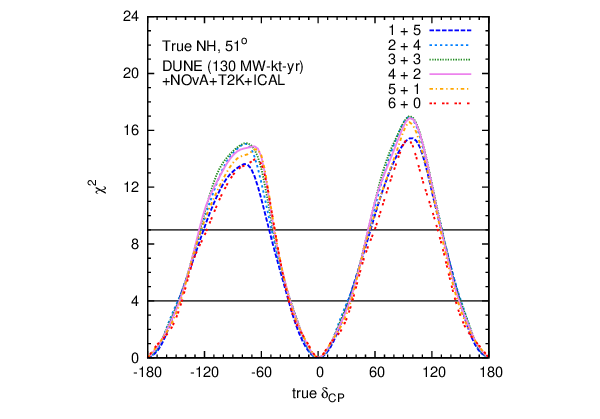, width=0.40\textwidth, bbllx=89, bblly=50, bburx=300, bbury=255,clip=}
 
\end{tabular}

\caption{Sensitivity for DUNE for various 
 combinations of neutrino and antineutrino run by itself (left panel) and in 
 conjunction with T2K, \nova\ and ICAL (right panel). The top/middle/bottom 
 row shows the sensitivity to hierarchy/octant/CP violation detection. 
 The total exposure has been divided into 6 equal parts and distributed
between neutrinos and antineutrinos. For example,
for hierarchy sensitivity, 6+0 corresponds to 44 MW-kt-yr in only neutrino; 
3+3 correspond to 22 MW-kt-yr in each neutrino and antineutrino mode. 
}
\label{fig:nuopt}
\end{figure*}

It is now well known that neutrino and antineutrino oscillation 
probabilities suffer from the same form of hierarchy-$\dcp$ degeneracy \cite{novat2k}. 
However, the octant-$\dcp$ degeneracy has the opposite form for neutrinos and 
antineutrinos~\cite{suprabhoctant,minakata_cp}.
Thus, inclusion of an 
antineutrino run helps in lifting 
this degeneracy
for most of the values of $\dcp$ \cite{suprabhlbnelbno}.
For  measurement of $\dcp$,
it has been  shown for T2K that   
the antineutrino run is required only for those true hierarchy-octant-$\dcp$
combination for which octant degeneracy is present \cite{Ghosh:2014zea}.   
Once this degeneracy is lifted by including 
some amount of antineutrino data, further antineutrino run does not 
help much in CP discovery; 
in fact it is then better to run with neutrinos to gain in 
statistics \cite{Ghosh:2014zea}.   
But this conclusion may change for a different baseline and matter effect. 
From Fig.~\ref{fig:beamprofile} we see that for \nova\
the oscillation peak does not coincide with the flux peak. 
Around the energy where the flux peaks,  
the probability spectra with $\dcp = \pm0, 180^\circ$ are not equidistant from the 
$\dcp = \pm 90^\circ$ spectra. For antineutrino mode the curves for 
$\pm 90^\circ$ switch position. Hence for neutrinos 
$\dcp = 0^\circ$ is closer to $\dcp = -90^\circ$ and $\dcp = 180^\circ$ is 
closer to $\dcp = 90^\circ$, while the opposite is true for 
antineutrinos. This gives a synergy and hence running in both neutrino 
and antineutrino modes can be helpful. 
For T2K the  energy where the flux peak occurs coincides with the oscillation peak. 
At this point the curves for $\dcp = 0, 180^\circ$ are equidistant from 
$\dcp = \pm 90^\circ$ and hence this synergy is not present. Thus, the role 
of antineutrino run is only to lift the octant degeneracy. 
The recent hint of $\dcp$ from T2K~\cite{t2k_dcphint} already 
gives us some evidence of the octant (see Table XXVIII in Ref.~\cite{t2k_dcphierocthint}, 
or Ref.~\cite{Ghosh:2014zea}). Moreover, \nova\ and DUNE 
will collect far more data with antineutrinos than T2K. 
Thus, the inclusion of antineutrino run at T2K does not make much 
difference to our results.
In the following we have varied the proportion  
of neutrino and antineutrino runs at DUNE to
ascertain what is the 
optimal combination. 
The adequate exposure is split into various combinations of neutrinos 
and antineutrinos -- 1/6$\ \nu$ + 5/6$\ \nubar$, 2/6$\ \nu$ + 4/6$\ \nubar$, ... 
6/6$\ \nu$ + 0/6$\ \nubar$. The intermediate configuration 3/6$\ \nu$ + 3/6$\ \nubar$ 
corresponds to the equal-run configuration used in the other sections. For 
convenience of notation, these configurations are referred to simply as 
1+5, etc., i.e. without appending the `/6'.
The results are shown in Fig.~\ref{fig:nuopt}. 

The top row of Fig.~\ref{fig:nuopt} shows the hierarchy sensitivity of 
DUNE for various 
combinations of neutrino and antineutrino run. Normal hierarchy and 
$\theta_{23}=39^\circ$ have been assumed as the true parameters.
For DUNE, we have chosen a total 
exposure of 44 MW-kt-yr which was found to be the adequate exposure in  
Section~\ref{sec:adequate} assuming equal neutrino and antineutrino runs. 
In the left panel, we see the results for DUNE alone. 
the figure shows that in the favourable region of $\dcp \in [-180^\circ,0]$ 
the best sensitivity comes from the combination 3+3 or 4+2. 
Although the statistics is more for neutrinos, the antineutrino run 
is required to remove the wrong-octant regions.
For NH, $\dcp \in [0,180^\circ]$ is the unfavourable region 
for hierarchy determination \cite{novat2k}, as is evident from the figure. 
In this region, we see that the results are worst for pure neutrino
run.   The best sensitivity comes for the case 5+1. 
This amount of antineutrino run is required to remove the octant degeneracy.
The higher proportion of neutrino run ensures better statistics.  
In the right panel, along with DUNE we have also combined data from \nova, 
T2K and ICAL. With the inclusion of these data the hierarchy sensitivity 
increases further and even in the unfavourable region $\chi^2=25$ sensitivity 
is possible with only neutrino run from DUNE. 
This is because \nova, which will run in antineutrino mode 
 for 3 years and the antineutrino component in the atmospheric 
neutrino flux at ICAL, will provide the necessary amount of information to 
 lift the parameter degeneracies that reduce hierarchy sensitivity. 
 Therefore, the best option for DUNE is to run only in neutrino mode, which 
 will have the added advantage of increased statistics. 
In the favourable region also the sensitivity 
is now better for 6+0 and 5+1  i.e. less amount of antineutrinos 
from DUNE is required because of the antineutrino information 
coming from \nova. Note that overall, the amount of antineutrino run 
depends on the value of $\dcp$. However combining information 
from all the experiments 4+2 seems to be the best option over the 
largest fraction of $\dcp$ values. 

In the middle row of Fig.~\ref{fig:nuopt}, we have shown the octant 
sensitivity of DUNE 
 alone (left panel) and in combination with the current experiments (right 
 panel). For DUNE we have used an exposure of 74 MW-kt-yr.
We have fixed the true hierarchy to be inverted, and 
 $\theta_{23}=39^\circ$ i.e. in the lower octant.
For this case the probability for neutrinos is maximum for 
$\dcp \sim -90^\circ$  and overlaps with the higher octant probabilities. 
Thus the octant sensitivity in neutrino channel is very poor. This the 
worst results for these values of $\dcp$ come from only neutrino runs. 
For antineutrino channel because of the flip in $\dcp$ the probability
for $\dcp = -90^\circ$ is well separated from those for HO. Therefore the 
octant sensitivity comes mainly from antineutrino channel \cite{suprabhoctant}. 
Thus, addition of antineutrino runs help in enhancing octant sensitivity.   
Therefore at $-90^\circ$ the best sensitivity is from 1+5 i.e $1/6^{th}$  
neutrino + $5/6^{th}$ antineutrino combination. 
On the other hand the neutrino probability is minimum 
for $\dcp = +90^\circ$ and  LO and therefore there is octant sensitivity 
in the neutrino channel. However since we are considering IH the 
antineutrino probabilities are enhanced due to matter effect and
for a broadband beam 
some sensitivity comes from the antineutrino channel also. 
Therefore there is slight increase in octant sensitivity 
by adding antineutrino data as can be seen. 
Overall, the best compromise is  seen to be reached for 
2+4 i.e $1/3^{rd}$ neutrino and 
$2/3^{rd}$ antineutrino combination, which gives the best results over 
the widest range of $\dcp$ values. 
Addition of \nova, T2K 
and ICAL data increases the octant sensitivity. The octant sensitivity 
is best for combinations having more antineutrinos. For $\dcp \sim +90^\circ$
all combinations give almost the same sensitivity.  
We have not presented the results for NH in this case.
For this case after adding T2K+\nova+ICAL to DUNE 
requires at least 4+2 to reach  $\chi^2=25$  for $\dcp \in [-180^\circ,0]$
while for $\dcp \in [0,180^\circ]$ 
the octant sensitivity almost crosses $\chi^2 = 25$ 
for all combinations of neutrino and 
antineutrino run.  
Therefore, the exact combination chosen does not make much difference to 
the final result.


The left and right panels of the bottom row in Fig.~\ref{fig:nuopt} show the 
ability  of DUNE (by itself, and in conjunction with the current generation 
of experiments, respectively) to detect CP violation.
Here the true hierarchy is NH and true $\theta_{23}$ is $51^\circ$. 
Although this true combination does not suffer from any octant degeneracy, 
we see in the left panel that 6+0 is not the best combination. 
This is due to the  synergy between neutrino-antineutrino runs
for larger baselines 
as discussed earlier. 
 In both cases, we find that 
 the best option is to run DUNE with antineutrinos for around a third of the 
 total exposure. 
On adding information from T2K and \nova, 
 we find great improvement in the CP sensitivity. From the right panel, we 
 see that the 
 range of $\dcp$ for which $\chi^2 = 9$ detection of CP is possible is almost 
 the same for most combinations of neutrino and antineutrino run. Therefore, 
 as in the case of octant determination, the exact choice of combination 
 is not very important.
 
  \section{Summary}
 
 The DUNE experiment at Fermilab has a promising physics potential. Its baseline 
 is long enough to see matter effects which will help it to break the 
 $\dcp$-related degeneracies and determine the neutrino mass hierarchy and the 
 octant of $\theta_{23}$. This experiment is also known to be good for 
 detecting CP violation in the neutrino sector. The current and upcoming 
 experiments T2K, \nova\ and ICAL@INO will also provide some indications 
 for the values of the unknown parameters. In this work, we have explored 
 the physics reach of DUNE, given the data that these other experiments will 
 collect. We have evaluated the adequate exposure for DUNE (in units of MW-kt-yr), 
 i.e. the minimum exposure for DUNE to determine the unknown parameters 
 in combination with the other experiments, for all values of the oscillation 
 parameters. The threshold for determination is taken to be $\chi^2=25$ for the 
 mass hierarchy and octant, and $\chi^2=9$ for detecting CP violation.
The results are summarized in Table ~\ref{tab:adequate}. 
We  find that adding information from \nova\ and T2K helps in reducing the 
exposure required by only DUNE for determination of all the three unknowns--
hierarchy, octant and $\dcp$. Adding ICAL data to this combination further   
help in achieving the same level of sensitivity with a reduction in exposure 
of DUNE (apart from $\dcp$). 
Thus the synergy between various experiments can be helpful in 
economizing the DUNE configuration. 
We have also  probed the role of the ND in improving the results 
 by reducing systematic errors. We have simulated events at the near and far 
 detectors and performed a correlated systematics analysis of both sets of events. 
 We find an improvement in the physics reach of DUNE when the ND is 
 included. We have also evaluated the drop in systematics because of the near
 detector. Our results are shown in Table~\ref{tab:systresults}. 
 
Further  we have checked the role of information from the lowest energy bins 
which are affected by the second oscillation maximum of the probability. 
We find that inclusion of these bins enhances the 
the hierarchy sensitivity since the hierarchy-$\dcp$ degeneracy has a complementary 
behaviour at the two oscillation maxima. Thus the increase in sensitivity  
is most significant in regions of parameter space where the degeneracies reduce 
the sensitivity. We find that the effect is more prominent when a greater 
exposure is required. For the combined analysis to reach $\chi^2=25$ one needs,
respectively, 44(56) MW-kt-yr including(excluding) the second oscillation maximum.
However for only DUNE the same sensitivity requires 436 MW-kt-yr but including the
second oscillation maximum the exposure is reduced to 212 MW-kt-yr to reach 
$\chi^2 = 25$.   
 
Finally, we have done an optimization study of the 
neutrino-antineutrino run for DUNE.  The amount of antineutrino run 
required depends on the true value of $\dcp$. 
It helps in achieving two objectives -- (i) reduction in 
octant degeneracy and (ii) synergy between neutrino and antineutrino data
for  octant and CP sensitivity. 
For a hierarchy determination using a total exposure 
of 44 MW-kt-yr
the optimal combination for only DUNE is (3+3) which corresponds to
22 MW-kt-yr in 
neutrino and antineutrino mode each, for $\dcp$ in the 
lower half-plane $[-180^\circ,0]$ and true NH-LO.   
For $\dcp$ in the upper half-plane ($[0,180^\circ]$) the  optimal ratio is 
$5/6^{th}$ of the total exposure 
in neutrinos and + $1/6^{th}$ of the total exposure in antineutrinos.   
Adding information from T2K, \nova\ and ICAL 
the best combination for DUNE is $2/3^{rd}$ neutrino + $1/3^{rd}$ antineutrino 
for $\dcp$ in the lower half-plane. 
In the upper half-plane, pure neutrino run gives the best sensitivity.  
In the latter case, the 
antineutrino component coming from \nova\ and 
ICAL helps in reducing the required antineutrino run from DUNE. 
For octant sensitivity the best result from the combined experiments 
comes from the proportion $(1/6^{th}+5/6^{th})$ except for 
$\dcp = +90^\circ$ where
all combinations
give almost the same sensitivity. For $\dcp$ all combinations give 
similar results when all data are added together, with equal neutrino and
antineutrino or $2/3^{rd}$ neutrino + $1/3^{rd}$ antineutrino combination 
faring slightly better.



To conclude, the DUNE experiment can measure mass hierarchy, octant and 
$\dcp$ with considerable precision. Inclusion of the data from the experiments 
like T2K, \nova\ and ICAL can help  DUNE to attain the same level of precision 
with a reduced exposure. Thus the synergistic aspects between different 
experiments can help in the planning of a more economized configuration for 
DUNE.

\begin{acknowledgements}
We would like to thank Daniel Cherdack, Raj Gandhi, Newton Nath and 
Robert Wilson for useful discussions. 
\end{acknowledgements}

\appendix
 \section{Computing the effect of the near detector on systematics}
 \label{app:syst}
 
 In this appendix, we discuss briefly the simple procedure that we have used to 
 combine results from the ND and FD, with correlated systematics. 
 This procedure is based on the method of pulls \cite{pulls_gg}. The ($1\sigma$)
 systematic errors are given by a set of numbers $\vec{\pi}$. 
 These errors can be normalization errors (which affect the scaling of events) 
 or tilt errors (which affect the energy dependence of the events). 
 The `experimental' 
 data $N^{det (ex)}_{i}$ are simulated using the `true' oscillation parameters 
 $\vec{p}_{ex}$, while the `theoretical' events $N^{det (th)}_{i}$ are generated 
 using the `test' oscillation parameters $\vec{p}_{th}$. The subscript $i$ here 
 runs over all the energy bins. The superscript $det$ can take values ND or FD. 
 The theoretical events get modified due to systematic errors as
 \begin{eqnarray*}
 M^{det (th)}_{i} (\vec{p}_{th}) & = & N^{det (th)}_{i} (\vec{p}_{th}) \Bigg[ 1 + {\displaystyle \sum_k} \xi_k \pi^k \\
 & + & {\displaystyle \sum_l} \xi_l \pi^l {\displaystyle \frac{E_i-E_{av}}{E_{max}-E_{min}} } \Bigg] ~,
 \end{eqnarray*}
 where the index $k(l)$ runs over the relevant normalization(tilt) 
 systematic errors for a given 
 experimental observable. All the pull variables $\{ \xi_j \}$ take values in the range 
 $(-3,3)$, so that the errors can vary from $-3\sigma$ to $+3\sigma$. Here, $E_i$ 
 is the mean energy of the $i^{\rm th}$ energy bin, $E_{min}$ and $E_{max}$ are the 
 limits of the full energy range, and $E_{av}$ is their average. 
 
 The Poissonian $\chi^2$ is calculated for each detector as
 \begin{eqnarray*}
  \chi^{2 \ det} (\vec{p}_{ex},\vec{p}_{th}; \{\xi_j\}) & = & {\displaystyle \sum_i} 2 \Bigg[ M^{det (th)}_{i}(\vec{p}_{th}) - N^{det (ex)}_{i}(\vec{p}_{ex}) \\
  & + & N^{det (ex)}_{i}(\vec{p}_{ex}) \ln \left( {\displaystyle \frac{N^{det (ex)}_{i}(\vec{p}_{ex})}{M^{det (th)}_{i}(\vec{p}_{th})} } \right) \Bigg] ~. 
 \end{eqnarray*}
The results from the two detector setups are then combined, along with a penalty 
for each source of systematic error. The final $\chi^2$ is then calculated 
by minimizing over all combinations of $\xi_j$, as 
\begin{align*}
 \chi^2 (\vec{p}_{ex},\vec{p}_{th}) & = & \min_{\{\xi_j\}} & \bigg[ \chi^{2 \ FD} (\vec{p}_{ex},\vec{p}_{th}; \{\xi_j\}) \\
 &&& + \chi^{2 \ ND} (\vec{p}_{ex},\vec{p}_{th}; \{\xi_j\}) \\
  &&& + {\displaystyle \sum_j} \xi_j^2 \bigg] \\
  & \equiv & \hspace{-2cm} \chi^2 ( {\rm{FD}} \oplus {\rm{ND}} ) ~. \span 
\end{align*}

Usually, for two experiments with uncorrelated systematics, the adding of penalties 
and minimizing over the pull variables is done independently, and the resulting 
$\chi^2$ values are added. In contrast, here we add the same pulls to both detector 
setups, and then minimize over the pull variables. This takes care of 
correlations between the systematic effects of the two setups.


\bibliographystyle{h-elsevier}
\bibliography{neutosc}   

%
%

\end{document}